\documentstyle [12pt] {article}

\catcode`@=12
\begin{document}
\thispagestyle{empty}
\begin{flushright}
UCRHEP-T117\\
MAD/PH/799\\
October 1993\\
\end{flushright}
\vspace{1.0in}
\begin{center}
{\Large \bf The Tau Neutrino as the\\
Lightest Supersymmetric Particle\\}
{\large $\rm (\nu_\tau = LSP)$\\}
\vspace{1.0in}
{\bf Ernest Ma\\}
\vspace {0.1in}
{\sl Department of Physics, University of California, Riverside, CA 92521\\}
and\\
{\sl Department of Physics, University of Wisconsin, Madison, WI 53706\\}
\vspace {1.0in}
\end{center}
\begin{abstract}\
Given the conservation of baryon number as well as electron and muon numbers,
all present data can be explained without the need of a conserved quantum
number
for the tau lepton and its neutrino.  A supersymmetric extension of the
standard model naturally allows for such a scenario.  One consequence is that
the tau neutrino becomes the lightest supersymmetric particle.
\end{abstract}
\vspace{0.4in}
\noindent \underline {~~~~~~~~~~~}\\
\noindent To appear in Proceedings of the Second Talinn Neutrino Symposium.
\newpage
\baselineskip 24pt

\noindent {\bf 1. Introduction}

It is usually thought that present experimental data indicate that
baryon number (B), electron number (L$^e$), muon number (L$^\mu$),
and tau number (L$^\tau$) should be taken as conserved quantum numbers,
at least to a very good approximation.  Actually, if we assume the
conservation of B, L$^e$, and L$^\mu$, then there is no need for
L$^\tau$.  For example, the fact that in neutrino interactions on nuclei,
$\tau$ leptons are not produced is explained by the conservation of L$^e$
and L$^\mu$.  Analogously, the fact that $\tau$ does not decay into an
antiproton + mesons requires only the conservation of B.

The reason that L$^\tau$ appears to be conserved is that given the standard
SU(2) $\times$ U(1) electroweak gauge model with its minimal particle content,
the conservation of B, L$^e$, L$^\mu$, and L$^\tau$ is automatic.  An
excursion beyond the standard model is necessary to see how L$^\tau$ may
be different from L$^e$ and L$^\mu$.\\

\noindent {\bf 2. Supersymmetric Standard Model}

Consider for example any supersymmetric extension of the standard model.  In a
notation where only left chiral superfields are counted, the quarks and leptons
transform under SU(3) $\times$ SU(2) $\times$ U(1) as follows:
\begin{equation}
Q \equiv (u,d)_{\rm L} \sim (3,2,1/6),~~u^c \sim (\overline 3, 1, -2/3),~~
d^c \sim (\overline 3, 1, 1/3),
\end{equation}
\begin{equation}
L \sim (\nu_\ell, \ell)_{\rm L} \sim (1,2, -1/2),~~~\ell^c \sim (1,1,1).
\end{equation}
The Higgs superfields are
\begin{equation}
\Phi_1 \sim (1,2,-1/2),~~~\Phi_2 \sim (1,2,1/2).
\end{equation}
The allowed terms in the superpotential are then
\begin{center}
\begin{math}
\Phi_1 \Phi_2,~~~\Phi_1 Q_i d_j^c,~~~\Phi_2 Q_i u_j^c,~~~\Phi_1 L_i \ell_i^c;
\end{math}
\end{center}
and
\begin{center}
\begin{math}
L_i \Phi_2,~~~L_i Q_j d_k^c,~~~L_i L_j \ell_k^c,~~~u_i^c d_j^c d_k^c.
\end{math}
\end{center}
Whereas the first four terms conserve B, L$^e$, L$^\mu$, and L$^\tau$, the
above four do not and they are usually just ignored.  What this amounts to is
the {\underline {imposition}} of B, L$^e$, L$^\mu$, and L$^\tau$ as conserved
quantum numbers.

Instead of forbidding all possible combinations of the last four terms, it is
just as natural to forbid some of them, resulting in the conservation of one
or two lepton numbers.  It has been shown\cite{1} that there are 5 $\times$ 3
ways of doing this, as given below.

\begin{center}
\begin{math}
\begin{array} {|c|c|c|c|} \hline
\rm Model & e & \mu & \tau \\
\hline
1 & (1,0) & (0,1) & (0,0) \\
2 & (1,0) & (0,1) & (1,1) \\
3 & 1 & -1 & 0 \\
4 & 1 & 1 & 0 \\
5 & 1 & 0 & 0 \\
\hline
\end{array}
\end{math}
\end{center}

\noindent Here the lepton-number assignments (L$^e$,L$^\mu$) are displayed
for models 1 and 2, and L for models 3-5.  For each model, there are of course
three possible variations resulting from the permutations of $e, \mu$, and
$\tau$.\\

\noindent {\bf 3. Zero Lepton Number for $\tau$}

If $\tau$ is to have zero lepton number, then Model 1 is the prototype.\cite{2}
The superpotential is given by
\begin{eqnarray}
W &=& \mu \Phi_1 \Phi_2 + h_i \Phi_1 L_i \ell_i^c + h_{ij}^u \Phi_2 Q_i u_j^c
+ h_{ij}^d \Phi_1 Q_i d_j^c \nonumber \\ &+& \mu' L_3 \Phi_2 + f_e L_3 L_1
\ell_1^c + f_\mu L_3 L_2 \ell_2^c + f_{ij} L_3 Q_i d_j^c,
\end{eqnarray}
where the coupling matrix $h_i$ has been chosen to be diagonal.  It reduces
to the conventional supersymmetric model in the limit $f_e, ~f_\mu, ~f_{ij}$,
and $\mu'$ go to zero.  The neutralino mass matrix spanning the gauginos
($-i\lambda_\gamma, -i\lambda_z$), the higgsinos ($\psi_1^0, ~\psi_2^0$),
and $\nu_3$ is then
\vspace{0.3in}
\begin{equation}
{\cal M} = \left[ \begin{array} {c@{\quad}c@{\quad}c@{\quad}c@{\quad}c}
c^2 M_1 + s^2 M_2 & sc(M_2-M_1) & 0 & 0 & 0 \\ sc(M_2-M_1) & s^2 M_1 +
c^2 M_2 & M_Z \cos \beta & -M_Z \sin \beta & 0 \\ 0 & M_Z \cos \beta &
0 & -\mu & 0 \\ 0 & -M_Z \sin \beta & -\mu & 0 & -\mu' \\ 0 & 0 & 0 &
-\mu' & 0 \end{array} \right],
\end{equation}

\vspace {0.2in}
\noindent where $c \equiv \cos \theta_W$, $s \equiv \sin \theta_W$,
$\tan \beta
\equiv v_2/v_1 = \langle \phi_2^0 \rangle / \langle \phi_1^0 \rangle$,
and $M_{1,2}$ are allowed gauge-invariant mass terms for the U(1) and
SU(2) gauginos.  As a result, the physical $\nu_\tau$
is mostly $\nu_3$, but there is a small admixture of the other states
so that it acquires a small see-saw Majorana mass
\begin{equation}
m_{\nu_\tau} \sim {\mu^{\prime 2} \over {2 \mu \tan \beta}}.
\end{equation}

\noindent For $\mu \sim 1$ TeV, tan $\beta \sim 10, ~\mu' \sim 10$
MeV, an interesting value of 5 eV is obtained for $m_{\nu_\tau}$.
\vspace{0.1in}

In the conventional supersymmetric model without the $\mu'$ term, the
smallest eigenvalue of the 4 $\times$ 4 neutralino mass matrix is that
of the lightest supersymmetric particle (LSP) and since R parity,
defined to be $(-1)^{\rm 2j+3B+L}$, is conserved in that case, the
LSP is absolutely stable.  This has very important implications for
the experimental search for supersymmetry because the LSP would be
a weakly interacting particle which leaves the detector unobserved
except that it carries away momentum and energy.  Here, it is
clear that $\nu_\tau$ is the LSP, so that any of the other
neutralinos such as the photino $\tilde \gamma$
({\it i.e.} $-i\lambda_\gamma$) would
decay through a scalar electron for example into $e^+ e^- \nu_\tau$
and might not remain stable within the detector in a typical
high-energy physics experiment.\\
\newpage
\noindent {\bf 4. Phenomenological Constraints and Implications}

The existence of the $L_3 Q_i d_j^c$ term is a source of tree-level
flavor-changing neutral currents.  The most stringent constraints
come from the neutral kaons.  The magnitude of the $\rm K_L - K_S$
mass difference implies
\begin{equation}
|f_{sd} f_{ds}^*|^{1 \over 2} < 10^{-4} \left( {m_{\tilde \nu_\tau}
\over {100~{\rm GeV}}} \right),
\end{equation}
and the $\rm K_L \rightarrow \mu^+ \mu^-$ rate implies
\begin{equation}
|f_\mu|^{1 \over 2} \left( |f_{sd}|^2 + |f_{ds}|^2 \right)^{1 \over 4}
< 5 \times 10^{-4} \left( {m_{\tilde \nu_\tau} \over {100~{\rm GeV}}}
\right).
\end{equation}
The possible deviation from zero of the Michel parameter $\eta$ in
$\mu \rightarrow e \nu \bar \nu$ decay implies
\begin{equation}
|{\rm Re}f_\mu f_e^*|^{1 \over 2} < 0.11 \left( {m_{\tilde \tau_{\rm L}}
\over {100~{\rm GeV}}} \right).
\end{equation}
The parameters $\rho$ and $\delta$ remain at 0.75 as in the standard
model.

Since the nonconservation of L$^\tau$ is only achieved here with the
intervention of supersymmetry, possible tests of this model will
likely involve the direct production of supersymmetric particles.
The best case is \cite{2}
\begin{center}
\begin{math}
p \bar p ~({\rm or}~ pp) \rightarrow \tau^\pm \tau^\pm +
2~{\rm quark~jets},
\end{math}
\end{center}
which involves (for the $\tau^+ \tau^+$ mode) the subprocesses
$u \bar d \rightarrow \tilde u_{\rm L}
\tilde d_{\rm R}^*$ through gluino exchange and their subsequent
decay into $d \tau^+$ and $\bar u \tau^+$ respectively.
The analogous reaction for an electron-proton collider is
\begin{center}
\begin{math}
ep \rightarrow \tau^- \tau^- + 1~{\rm quark~jet + missing~energy},
\end{math}
\end{center}
which proceeds via the production of $\tilde e_{\rm R}
\tilde d_{\rm R}$
and their subsequent decay into $\nu_e \tau^-$ and $u \tau^-$
respectively.
\newpage
\noindent {\bf 5. Conclusion}

If supersymmetry exists, the issue of lepton number conservation will
have to be reexamined experimentally.  The conventional wisdom that
L is conserved additively (or multiplicatively if neutrinos have
Majorana masses) may not be correct.  An interesting possibility
would be that $\tau$ is actually a nonleptonic superparticle and
\begin{center}
\begin{math}
\begin{array} {|c|} \hline
\nu_\tau = {\rm LSP}\\
\hline
\end{array}
\end{math}
\end{center}
\vspace{1.3in}
\begin{center}
{ACKNOWLEDGEMENT}
\end{center}

I thank Professors Ots and Tammelo for their great hospitality and
a very stimulating meeting.  This work was supported in part by the
U.S. Department of Energy under Contract No. DE-AT03-87ER40327.

\bibliographystyle{unsrt}

\begin{thebibliography} {99}
\bibitem{1} E. Ma and D. Ng, {\em Phys. Rev.} {\bf D41}, 1005 (1990).
\bibitem{2} E. Ma and P. Roy, {\em Phy. Rev.} {\bf D41}, 988 (1990).

\end{thebibliography}

\end{document}